\documentclass[oldversion]{aa}
\usepackage{graphicx}
\usepackage{natbib}
\usepackage{txfonts}
\bibpunct{(}{)}{;}{a}{}{,}

\begin{document}

%------the header------------------------------------
\title{Finding pulsars with LOFAR}

\author{
Joeri van Leeuwen \inst{1}
\and 
Ben Stappers      \inst{2}
}

\institute{ 
  Stichting ASTRON, PO Box 2, 7990 AA Dwingeloo, The Netherlands, {\it leeuwen@astron.nl}
\and
  Jodrell Bank Centre for Astrophysics, School of Physics and Astronomy, University of Manchester, Manchester M13 9PL, UK
}

\date{Received / Accepted}

%------the abstract----------------------------------
\abstract{ We investigate the number and type of pulsars that will be
  discovered with the low-frequency radio telescope LOFAR. We consider
  different search strategies for the Galaxy, for globular clusters
  and for other galaxies. We show that a 25-day all-sky Galactic
  survey can find approximately 900 new pulsars, probing the local
  pulsar population to a deep luminosity limit. For targets of smaller
  angular size such as globular clusters and galaxies many LOFAR
  stations can be combined coherently, to make use of the full
  sensitivity. Searches of nearby northern-sky globular clusters can
  find new low luminosity millisecond pulsars. Giant pulses
  from Crab-like extragalactic pulsars can be detected out to over a
  Mpc.  }


\maketitle

%------the introduction------------------------------
\section{Introduction}

Since the discovery of the first four pulsars with the Cambridge radio
telescope \citep{hbp+68}, an ongoing evolution of telescope systems
has doubled the number of known radio pulsars roughly every 4 years:
An evolution from a large flat receiver with a fixed beam on the sky
(the original Cambridge radio telescope) to focusing dishes
\citep[Arecibo --][]{ht75b}, often steerable (Green Bank Telescope),
on both hemispheres \citep[the Parkes telescope --][]{mlc+01}, with
large bandwidths and multiple simultaneously usable receivers for
wider fields of view (Parkes, Arecibo). The types of pulsars
discovered have changed accordingly, from slow, bright, single and
nearby pulsars (the original four) to fast (young and millisecond
pulsars), far-away (globular clusters) or dim pulsars, some of which are in binaries.

The next step in radio telescope evolution will be the use of large
numbers of low-cost receivers that are combined
interferometrically. These telescopes, the Allen Telescope Array
\citep{bow07b}, LOFAR \citep{rot03}, MeerKAT \citep{jon07}, ASKAP \citep{jtb+08} and the SKA
\citep{kram04}, create new possibilities for pulsar research.

In this paper, we investigate the prospects of finding radio pulsars
with LOFAR, the LOw Frequency ARray. We outline and compare strategies
for targeting normal and millisecond pulsars (MSPs), both in the disk
and globular clusters of our Galaxy, and in other galaxies.

%----------------------------------------------------
\section{LOFAR - The LOw Frequency ARray}
\label{sec:LOFAR}

With the first test station operational and the first pulsars detected
LOFAR is on track to start operation in 2010. We have evaluated and
simulated the LOFAR reference configuration for pulsar searches, and
will describe that configuration in some detail below.

%% \begin{figure}[b]
%%   \includegraphics[]{0329.eps}
%%   \caption{
%%     {\em Include this picture? ref S+K 07}
%% }
%%   \label{img:0329}
%% \end{figure}

%% REFS BELOW:
%% http://www.lofar.org/workshop/23Apr07_Monday02/LOFARWorkshop_Apr07_JaapBregman.pdf p.5
%% http://www.lofar.org/workshop/23Apr07_Monday01/LOFARWorkshop_Apr07_CorinaVogt.pdf p.11

Using two different types of dipoles, LOFAR can observe in a low and a
high band that range from 30-80\,MHz and 110-240\,MHz
respectively. The sensitivity using the high-band antenna (HBA) is
several times that of the low-band antenna (LBA) although their survey
speeds are similar due to the larger LBA field of view. The low band
is expected to be an exciting new window for exploring radio pulsar
behavior \citep[cf.\ ][ for an overview of the possibilities for
  emission physics and interstellar medium studies]{slk+07}, but the
impact on a pulsar survey of some of the smearing effects further
discussed is so strong in the LBA band, that we will only discuss the
HBA half of LOFAR in this paper.

The basic collecting elements are the individual dual-polarization
dipoles; each 4x4 set of these dipoles is combined in an analog
beamformer and forms an antenna 'tile'.  Tiles are grouped together in
stations; stations farther from the array center are larger than inner
stations. In the core, HBA stations are grouped in pairs. The
innermost 12 stations are 24 tiles each and are packed tightly in a
'superstation'. Spread over the 2-km core there are 24 more HBA
stations of 24 tiles each (making for 36 core stations in total). These
core stations are 32 meters in diameter, but are tapered to an about
30 meter effective diameter to reduce sidelobes. Next there are 18 Dutch
'remote' stations that are outside the core and consist of 48 tiles
each, while the $\sim$8 international stations that are spread over
Europe use 96 HBA tiles.

%% http://www.lofar.org/workshop/23Apr07_Monday02/LOFARWorkshop_Apr07_JaapBregman.pdf#2 p.5
%%  + /operations/lib/exe/fetch.php?id=public%3Amsss_meeting2&cache=cache&media=public:msss2_gdb.pdf

\begin{figure}[b]
  \includegraphics[]{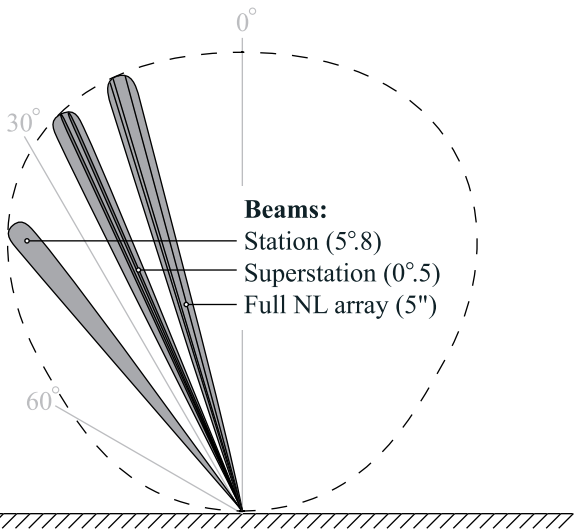}
  \caption{
  Illustration of station, superstation and full-array beam
  patterns at 120\,MHz. The zenith is at $0^\circ$. The dashed
  envelope is the tile response as shown 
  in Fig.\ \ref{img:a_eff}. The 
  gains of the different beams are here scaled to be the same. Three
  stations beams are shown, up to eight can be formed at a given
  time.
} 
  \label{img:beams}
\end{figure}

At each station the tiles are combined to form up to 8 independently
steerable 'station beams'.  With the cumulative data rate out of the
station being the limiting factor, the product of the number of beams
times their bandwidth cannot exceed 32\,MHz (potentially 48\,MHz),
e.g.: a station beam set up can range from having a single
full-bandwidth station beam to 8 independent beams of 4\,MHz
each. Station beams are subdivided in 195\,kHz channels, and sent to
the central processor (CEP) supercomputer for correlation, addition
and/or different types of beam forming. As illustrated in
Fig.\ \ref{img:beams} CEP can further combine station beams to form
$\sim$128 16-bit full-polarisation tied-array beams for the
superstation, the core and/or the entire array. That number of formed
beams is limited by the maximum CEP output rate, currently 50\,Gigabits/second.

\begin{figure}[t]
  \includegraphics[]{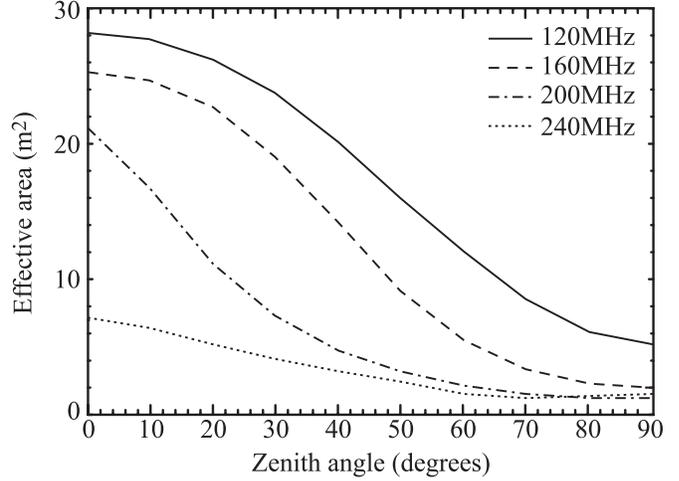}
  \caption{
    Simulated effective area per antenna tile versus zenith angle, for 4
  different observing frequencies. From \cite{wil07}.
}
  \label{img:a_eff}
\end{figure}

%% order ~Tbps, exact number unclear.
%% 12b x 100 M x 32 * 100 = 10^4 M * 400 = 4 Tbps
%% http://www.skatelescope.org/SKAmeeting_Paris/pdf/LOFAR20060905ska.pdf

The high-band antennas operate in a 110--240\,MHz frequency range and
are spaced to optimize sensitivity for the low end. They are maximally
sensitive toward the zenith. The fall-off of effective area $A_{eff}$
with zenith angle and observing frequency has been well characterized
as shown in Fig.\ \ref{img:a_eff} \citep[after][]{wil07} and we will
also describe it qualitatively here: at its maximum at 120\,MHz, the
effective area toward zenith is 28\,m$^2$ per antenna tile, halving at a
zenith angle of $\sim$50$^\circ$. Toward higher frequencies both
quantities decrease until at 240\,MHz the maximum effective area is
8\,m$^2$ per antenna tile, dropping to half that at $\sim$30$^\circ$ away
from zenith. For the N$_{station}\,$=\,36 stations of the compact core
at 120\,MHz, with their N$_{tiles}\,$=\,24 tiles each, this effective
area translates into a theoretical maximum core gain G$_{max, core}$
of

\begin{equation}
   G_{max, core} = \frac{28\,{\textrm m^2} ~ N_{stations} ~ N_{tiles} }{2\,k} 
             \times 10^{-26} 
             \frac{ \textrm{W\,s}     }
                  { \textrm{Jy}\,\textrm{m}^2  }
                  = 8.8 \, \textrm{K/Jy}
\label{eq:g}
\end{equation}

where {\em k} is Boltzmann's constant in W s m$^2$. Finally, the noise
temperature T$_{antenna}$ of the HBA is expected to be 140\,K
at 120\,MHz and around 180\,K in the upper half of the band
\citep{gun07}.

%----------------------------------------------------
\section{Pulsar searches}
\subsection{All-sky surveys}

Although a pulsar survey generally addresses several science questions in
parallel, it can be optimized for a specific goal: by using
short integration times potential acceleration smearing is kept to a
minimum to optimize for finding millisecond pulsars; when aiming to maximise
the total number of new pulsars found, a survey should
generally focus on the Galactic plane; while for a representative
understanding of the local population an all-sky survey that is
equally sensitive in all directions is optimal. Here we will focus on
such an all-sky survey.

Below we first investigate different beam forming scenarios. These we
compare assuming use of a single station beam, at full bandwidth, at the
lowest observing frequency of 120\,MHz. We investigate frequency and
band width dependencies in more detail in subsection \ref{sec:of}.

\subsubsection{Beam forming}
 \label{sec:bf}

To first order an all-sky pulsar survey is optimized to reach best
sensitivity per given overall time. For that one should
minimise system noise while maximising collecting area, bandwidth, and
integration time per pointing. 

For a given survey time and sky area, maximising the time per pointing
is equivalent to maximising the instantaneous field of view. This can be
done by forming multiple simultaneous beams; while for
sparse telescopes like LOFAR, where receivers are spread out over a
large area, one can also add stations incoherently instead of
coherently. This increases the field of view at the cost of
increasing uncorrelated noise over correlated signal
\citep[cf.~][]{bac99}. In the incoherent case the beam on the sky will
be larger, but in the coherent case the system is more sensitive.

From the above we can construct the following survey figure of
merit FoM for a given observing frequency \citep[see][for related definitions]{cor02b, sks+09}:
\begin{equation}
    FoM = 10^{-3} \left( \frac{A}{T_{sys}} \right) \left( \frac{N_{beams} ~ \Omega ~ B} {1|N_{stations}} \right)^{1/2} 
\label{eq:fom} 
\end{equation}

where the factor $10^{-3}$ scales the reference FoM to be 1.0;
$A/T_{sys}$ is the ratio of effective area and system noise equivalent
temperature, averaged over the field of view, a ratio indicating
telescope gain.  $N_{beams}$ is the number of simultaneous beams
formed and $\Omega$ denotes the field of view as derived from the beam
size. Here and below we estimate all beam sizes by their full width at
half maximum (FWHM) as $1.22\frac{\lambda}{D}\frac{360}{2\pi}$\,deg with
$\lambda$ the observing wavelength, and $D$ the diameter of the
element over which the beam is formed (e.g.: a station, the core, or
the full array). Then $\Omega = \pi(\frac{FWHM}{2})^2$\,deg$^2$. $B$ is the
observing bandwidth; the right-hand denominator is 1 for
coherent addition, or $N_{stations}$ when that number of stations is
added incoherently.  This FoM is inversely proportional to the minimum
detectable flux $S_{min}$. When comparing between setups, the
  ratio of survey speeds -- the time needed to reach a given
  sensitivity over the 
  whole sky -- is FoM$^2$. In Table \ref{tab:fom} we list the
FoM for three survey set ups discussed below.

%------foms------------------------------
\begin{table*}[]
  \centering
  \begin{tabular}%
  {lcccccc}
  \hline
  Type               & ${A}/{T_{sys}}$ (m$^2$/K) 
                                       & $N_{beams}$ & $\Omega$ (deg$^2$)
                                                             & $B$ (MHz)
                                                                    &  $1|N_{stations}$ 
                                                                          & $FoM$ \\
  \hline  \hline                                				    
  Full Incoherent        &    260          &1          & 27      &   32 &  54 &  1.0 \\ 
  Core Coherent          &    170          &128        & 0.006   &   32 &   1 &  0.8 \\
  Superstation Coherent  &     40          &128        & 0.23    &   32 &   1 &  1.2 \\
  \hline
  \end{tabular}

  \caption{ Figure of merit FoM per Eq.\ \ref{eq:fom} for the three different survey set ups
    as described in the text: Full Incoherent, Core Coherent and
    Superstation Coherent. All parameters derived using the lowest
    observing frequency, 120\,MHz. In the first column,
    $A/T_{sys}$, we use 54, 36 and 12 stations respectively,
    of 24 tiles at 28\,m$^2$, and 140\,K, while for the {\em Superstation
    Coherent} case we also multiply by 0.70 to correct for the lower
    sensitivity at the station beam edge.  \label{tab:fom}
 }
\end{table*}

To outline the various trade-offs involved with these survey setups,
we first compare scenarios for coherent versus incoherent addition for
the LOFAR core only, and not the total array -- given the sparseness
of the distribution of remote stations, coherent addition of the
entire array would result in an extremely limited field of view. 

In the {\em Core Coherent} scenario all 36 stations in the 2\,km
diameter $D_{core}$ compact core are combined coherently. In this case
$\Omega\,$=$\,0.0060\,$deg$^2$ and the gain is as per Eq. \ref{eq:g}.
For comparison, each of the individual roughly $D_{station}\,$=
30-meter core stations forms a beam on the sky of 27\,deg$^2$.  If
added incoherently, the resulting beam is as large as for an
individual core station (cf.\ Fig.\ \ref{img:beams}) but at a factor
$\sqrt{N_{stations}}=6$ decreased sensitivity compared to the coherent
case.  The factor
$(\frac{D_{core}}{D_{station}})^2\,$=$\,(\frac{2000\,\textrm{\tiny
    m}}{30\,\textrm{\tiny m}})^2\,$=$\,4.4 \times 10^3$ increase in
beam area allows for longer integrations by the same factor. For
synthesis telescopes that form a single coherent 'tied-array' beam,
like the Westerbork Synthesis Radio Telescope or the Very Large Array,
such a factor would well describe the trade-off; but as the LOFAR CEP
can output $N_{beams}$\,=\,$\sim$128 coherently added core beams
simultaneously versus only a single incoherently added core beam, the
factor $F_t$ difference in integration times is
$F_t\,$=$\,(\frac{D_{core}}{D_{station}})^2/N_{beams}\,$=$\,(\frac{2000\,\textrm{\tiny
    m}}{30\,\textrm{\tiny m}})^2/128\,$=$\,35$.  Thus, in this
comparison the sensitivity decrease by
$\sqrt{N_{stations}}\,$=$\,\sqrt{36}$ of adding stations {\em
  incoherently} and the sensitivity increase by
$\sqrt{F_t}\,$=$\,\sqrt{35}$ provided by longer integrations
practically cancel out: the FoM for coherent or incoherent addition of
just the core is very similar.

Although for the {\em coherent} sum the addition of remote stations
significantly reduces the field of view, for the incoherent case the
entire array can be added without field-of-view penalty. Physically
the remote stations are twice the size of the core stations but they
can be tapered to the size of the core stations to produce a similar
field of view. Adding these 18 remote stations in the above equation
produces a FoM for the {\em Full Incoherent} scenario of 1.0; that is
$(F_t \, N_{stations, full})^{1/2}/ N_{stations, core}\,$=$\,(35
\times 54)^{1/2}/36\,$=$\,1.2$ times higher than FoM for the {\em
  Coherent Core} case which is 0.8 (Table \ref{tab:fom}). Yet (only)
for the {\em Coherent Core} survey producing more simultaneous beams
increases the FoM; if 50\% more beams could be formed these two
scenarios would be equally sensitive.

Incoherent addition may offer slightly better sensitivity because it
allows for long integration times. If however these long integration
times are a significant part of a potential pulsar binary orbit
\citep[cf.\ 1-hr integrations in our reference incoherent survey,
  defined below, versus the 2.4-hr binary period of
  PSR~J0737-3039][]{bdp+03}, the apparent change in pulse period due
to the acceleration through the orbit decreases the effectiveness of
periodicity searches. At a computational cost different types of
acceleration searches can mitigate this decrease in sensitivity
\citep{jk91,rce03}. A coherent survey is less prone to this
acceleration smearing: to reach the same $S_{min}$ as the incoherent
reference survey with 1-hr pointings, the coherent addition survey
integration time is down to 3 minutes. From an efficiency point of
view short pointings are also more robust to system errors such as
data glitches, and to impulsive radio interference.  In contrast
  to searching a single beam for an incoherent-addition
  survey, handling $\sim$128 simultaneous data streams, each at high
  time resolution and totaling 50\,Gigabits/second is a significant
  computational and data-handling task (\citealt{bac99};
  cf.\ \citealt{sks+09} for the SKA).

In a different coherent survey,  {\em Superstation Coherent}, 128
tied-array beams 
from just the superstation (cf.\ section \ref{sec:LOFAR})
tile out the complete station beam. Compared to coherently adding the
entire core, the superstation beams have larger field of view but are
less sensitive. In a grid of 128 of these wider superstation beams
many will be close to the less sensitive edges of the station beam,
causing for some decrease in sensitivity compared to the narrow
 {\em Coherent Core} beams. For $A/T_{sys}$ in Table \ref{tab:fom} we
hence use the average sensitivity out to FWHM for an assumed Gaussian
station beam, which is 70\% of the sensitivity for the center of the
beam. Yet the relatively large field of view caused by the dense packing
of the superstation offsets the limited sensitivity of the lower
number of stations now used, producing a FoM of 1.2, compared to 1.0
for the incoherent survey discussed above. 
Given identical overall survey duration, the integration times for
this coherent superstation and the
incoherent full array are similar, so there is no reduction in acceleration
smearing. Inherent to coherent-addition surveys, the computational
task of searching through 128 beams simultaneously and in real-time
remains.

Beam forming scenarios have an impact on how accurately sources
can be located; this we will address in more detail in
section \ref{sec:disc} where we discuss follow up.

\subsubsection{Observing frequency}
 \label{sec:of}
Determining the optimum observing frequency involves mapping out the
trade-off between beam size, source and background brightness, and
pulse smearing from dispersion and scattering. Some telescope and
pulsar behavior favors low-frequency observing: when moving up from
the lowest frequencies the beam size decreases as $\nu^{-2}$ which
limits integration time per pointing in a fixed-time all-sky survey;
the effective collecting area falls off with frequency; and the
intrinsic pulsar brightness decreases with frequency as $\nu^{-1.5}$
on average \citep{mms00}. Other circumstances are more favorable at
higher frequencies: the background sky noise decreases steeply as
$\nu^{-2.6}$ \citep{lmo+87} and smearing from interstellar dispersion
and scattering, which makes the pulsar periodicity harder to detect,
is less.  Much of the effect of dispersion can be removed by searching
over many ($\sim$$10^4$) incoherently dedispersed trial DMs
\citep{lk05}.  No such removal process exists for scatter broadening,
where the pulsar emission takes different paths through the
interstellar medium with different travel times, resulting in an
observed pulse profile that is significantly smeared out. Scattering
smearing time increases sharply with decreasing observing frequency
(as $\nu^{-4.5}$, \citealt{rmdm+97}; or alternatively $\nu^{-3.9}$,
\citealt{bcc+04}) and with increasing distance to the pulsar
\citep{tc93}.

\begin{figure}[]
  \includegraphics[]{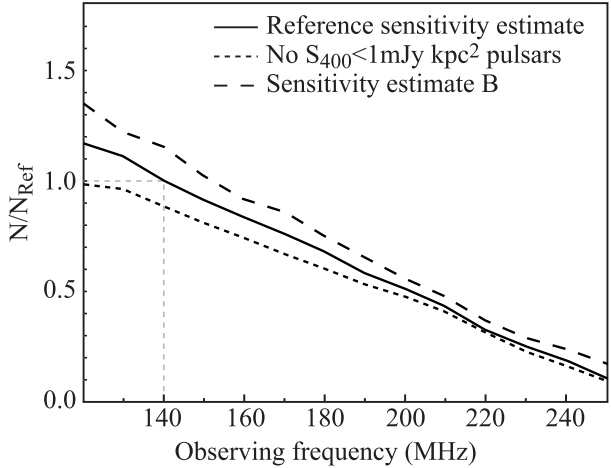} \caption{ Number of pulsars
    detected in simulated surveys at different frequencies and
    sensitivity estimates, plotted as a ratio over the productivity of
    our reference 25-day all-sky survey of 1-hr pointings at
    140\,MHz. Towards lower frequencies the yield goes up, mostly
    caused by the increasing effective area as the wavelength
    approaches the telescope's optimum at twice the antenna spacing of
    1.25\,m; equally important is the increasing beam size which
    allows for longer integrations for the same total survey time.  }
\label{img:n-freq}
\end{figure}

Although the sky background and scatter broadening increase towards
lower frequency we find in our simulations, further described below,
that the total survey productivity, if defined as the total number of
pulsars detected (Fig.\ \ref{img:n-freq}), is mainly determined by the
effective collecting area and the beam size. These peak towards the
lower edge of the band, leading us to conclude that the survey is most
efficient at lowest frequencies. When using a single full-bandwidth
32\,MHz station beam this means a central frequency of 140\,MHz, from here on
our reference frequency. As can be seen from Eq. \ref{eq:fom},
bandwidth can be freely traded for beams with no impact on the FoM. Splitting up
the available bandwidth over multiple independently pointing stations
beams (keeping beams-bandwidth product equal to 32\,MHz as outlined in
Section \ref{sec:LOFAR}) and moving each beam to a lower, more
sensitive observing frequency may therefore further increase the
overall survey output. This does increase integration time per
pointing, which may be only beneficial in the  {\em Core Coherent}
scenario where short default integration times could limit sensitivity
to intermittent pulsars.

\subsubsection{Sensitivity}
The minimum detectable flux S$_{min}$ depends on the
signal-to-noise ratio at which one accepts a signal as real (SNR), the
system temperature
(T$_{sys}$=T$_{rec}$+T$_{sky}$) as it varies
over the sky, the number of polarisations (N$_{pol}$), the
bandwidth (B), the integration time (t), the pulsar period (P) and pulse width (W)
and the zenith-angle dependent gain G(z).  Compared to the theoretical
gain the real-life gain is always less; although in Eq. \ref{eq:g} we
already use the simulated {\it effective} area and take aperture
efficiency into account there, we apply a conservative factor 0.66 to
estimate the real-life gain for our simulations below.
% Reference is
%  54 stations (18*2halves in core + 18 NL stations tapered) 
%  SNR=10
%  gain = 0.66 G_theory
%  REFB = 0.8 G_theory
Regarding the impact of RFI on the amount of usable bandwidth, testing
has shown the radio-interference environment to be relatively clear,
and RFI is significantly reduced by the 12-bit digitisation and the
many-element interferometer design, especially when elements are
combined coherently. As the total bandwidth can furthermore be quickly
split up and spaced to avoid channels contaminated by RFI, we expect
that the entire specified 32\,MHz band (potentially 48\,MHz) will be
usable, but below we shall conservatively use 80\% of 32\,MHz. The
potential 1.5-fold increase in usable bandwidth from 32 to 48\,MHz
would increase sensitivity by $\sqrt{48/32}$.  As S$_{min}$ depends on
zenith distance and sky noise it varies per pointing, but after
\citet{dtws85} we can estimate a typical value for a 1-minute pointing
towards the zenith using a coherently formed core beam, for a pulsar
with a 10\% duty cycle:

\begin{eqnarray}
S_{min} & = & {SNR} \left(\frac{T_{sys}}{0.66 \, G_{max} \, (N_{pol} ~ 0.8 B ~ t)^{\frac{1}{2}}   } \right)
                   \left(\frac{W}{P-W}\right)^{\frac{1}{2}} \nonumber
                   \\
       & = & 10 \left(\frac{1.0 \times 10^3\,\textrm{K}}{0.66 \times
                     8.8 \frac{\textrm{\tiny K}}{\,\textrm{\tiny Jy}} ~
                     (2 \times 0.8 \times 32\,\textrm{MHz} \times 60\,\textrm{s})^\frac{1}{2}  }\right)
    \left(\frac{1}{3}\right) \nonumber \\
& = & 10.3\,\textrm{mJy}
\label{eq:smin}
\end{eqnarray}

For comparison the $S_{min}$ for incoherently added pointings for our reference
survey described below (1-hr integrations, 54 stations added) comes
to 5.8\,mJy.

%------surveys------------------------------
\begin{table}[b]
  \centering
  \begin{tabular}%
%  {|l|l|r|r|r|}
  {llrrr}
  \hline
  Survey                 & Year      & N$_\textrm{real}$
                        		     & N$_\textrm{sim}$
                        				    & $\nu$ (MHz) \\
  \hline  \hline	        				    
  Jodrell                & 1972      & 51    &   20 $\pm$   2 &  408 \\
  UMass-Arecibo          & 1974      & 50    &   39 $\pm$   4 &  430 \\
  MolongloII             & 1978      & 224   &  227 $\pm$   7 &  408 \\
  UMass-NRAO             & 1978      & 50    &   54 $\pm$   6 &  400 \\
  ParkesII               & 1991-1994 & 298   &  335 $\pm$  10 &  400 \\
  Cambridge 80MHz        & 1993-1994 &  20   &   27 $\pm$   5 &   82 \\
  Parkes Multibeam       & 1997-2000 &   987 &  801 $\pm$  41 & 1374 \\
  LOFAR                  & 2010-2012 & --    & 1100 $\pm$ 100 &  140 \\
  \hline
  \end{tabular}
  % these are total numbers, not new numbers (cf. Cambridge 80MHz)
  \caption{Name, year, number of detected pulsars (N$_\textrm{real}$),
  number of simulated detected pulsars
  (N$_\textrm{sim}$) and frequency observed at ($\nu$) for the eight
  surveys simulated. The error on N$_\textrm{sim}$ is the standard
  variation in the outcomes when simulating the model described in the
  text with different random number seeds. The N$_\textrm{real}$ data
  is taken from \citet{dls77,ht75b,mlt+78,dth78,mld+96,stw98} and \citet{mhth05}
  respectively. The LOFAR numbers are for the incoherent 25-day reference design
  described in the text.}

  \label{tab:surveys}
\end{table}

\subsubsection{Simulations}
To determine the number and type of pulsars LOFAR can find, we have
used the above characteristics to model the detection of a large
number of simulated normal pulsars. For this, we have used a
population synthesis code that simulates the birth, evolution, death
and possible detection of radio pulsars \citep[for details see
][]{bwhv92,hbwv97a,lv04}. We do not simulate {\em millisecond}
pulsars, as a realistic treatment of the survey selection effects
associated with their binary history would be essential but is beyond
the scope of this work. For our simulated pulsars we draw birth
  velocities from the \citet{ll94} distribution with its $450 \pm
  90$\,km\,s$^{-1}$ mean velocity. Initial positions are simulated as
  in \citet{hbwv97a}: heights above the Galactic plane are randomly
  selected from an exponential distribution with a scale height of
  60\,pc; for the initial distribution in galactocentric radius we
use their Eq.\ 4, modified to scale {\em per unit area}:
\begin{equation}
  \centering
  \textrm{p(R)}\,\textrm{dR}= \frac{\textrm{R}}{\textrm{R}_w} ~
  \exp\left(-\frac{\textrm{R}}{\textrm{R}_w}\right) \, \textrm{dR}
\label{eq:r} 
\end{equation}
where scale length $R_w$ is 5\,kpc.  This modified distribution does
not peak as strongly toward the Galactic centre; for the local
population there is no significant change but surveys that sample the
inner Galaxy are better reproduced. We next calculate the 3D orbits
through the Galaxy. We evolve magnetic fields and periods, determine
whether the pulsar is still above the death-line. We estimate the
luminosity at 400\,MHz (per \citealt{hbwv97a}, who follow
\citealt{no90}). With a spectral index drawn from a Gaussian
distribution of width 0.76 around -1.5 \citep[][takes luminosity
  turn-overs into account]{mms00} we scale this luminosity to the
observing frequency.

Using the simulated position we look up the sky background temperature
\citep{hssw82} and combined with the distance we also estimate the dispersion
measure and the scatter broadening from \citet{tc93}. Given that both sky
background and scatter broadening time increase toward lower observing
frequencies we next scale these using power laws with spectral index
$-$2.6 and $-$4.4 respectively \citep{lmo+87,tc93}.

For each of the 8 pulsar surveys in Table \ref{tab:surveys} we model
the sensitivity function versus period, dispersion and scattering
smearing, and sky position. We also model the decrease of pulse width
and hence detectability with period. By comparing the
simulated and real pulsar samples for the first four of the above
surveys \citet{hbwv97a} determined the most probable underlying
initial model parameters.

We use this best model to determine the yield of a future survey with
LOFAR. As tests, we also model the second 81.5-MHz Cambridge survey
\citep{stw98} to check the validity of our extrapolations to lower
frequencies and the Parkes Multibeam pulsar survey \citep[][hereafter
PMB]{mlc+01, lfl+06} because of its superior statistics. These tests we
describe in some more detail in the two paragraphs below.

\begin{figure}[]
  \includegraphics[]{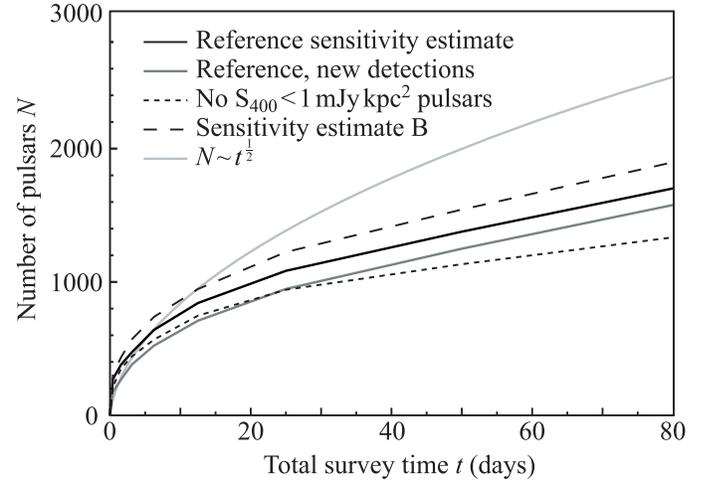}
  \caption{
   Simulated number of pulsars detected in an all-sky survey, versus total survey time
   in days. A survey time of 25 days corresponds to 
   1-hr pointings for incoherently added stations beams. 
}
  \label{img:n-vs-t}
\end{figure}

For each simulated pulsar, we model if it would be detected by the
Cambridge survey given its sky position, dispersion and scatter
smearing, and luminosity. We estimate the minimum detectable flux from
the $S_{min}$ versus DM and period plot in \citet{stw98}, their
Fig.\ 3.  That figure does not take scattering into account, although
for these simulations one realistically should.  For each simulated
pulsar we therefore determine the 'effective' (higher) DM-value that
would produce the same smearing as the actual dispersion and
scattering do combined.  We next determine the sensitivity for the
period and that 'effective' DM. The sensitivity thus estimated is
valid for the survey's median sky background noise of 2 Jy, so we next
scale it using the actual sky noise for the position of the simulated
pulsar. Finally we multiply the sensitivity by the
declination-dependent power response \citep[Fig 1a.,][]{stw98} for the
sky position and check if the simulated pulsar is bright enough to be
detected. With 27 $\pm$ 5 simulated pulsars found, our model
reproduces the actual tally of 20 pulsars quite well, validating the
luminosity, sky-noise and scattering extrapolations to lower
frequencies that we also use for the LOFAR predictions.

\begin{figure*}[t]
   \includegraphics[width=\textwidth]{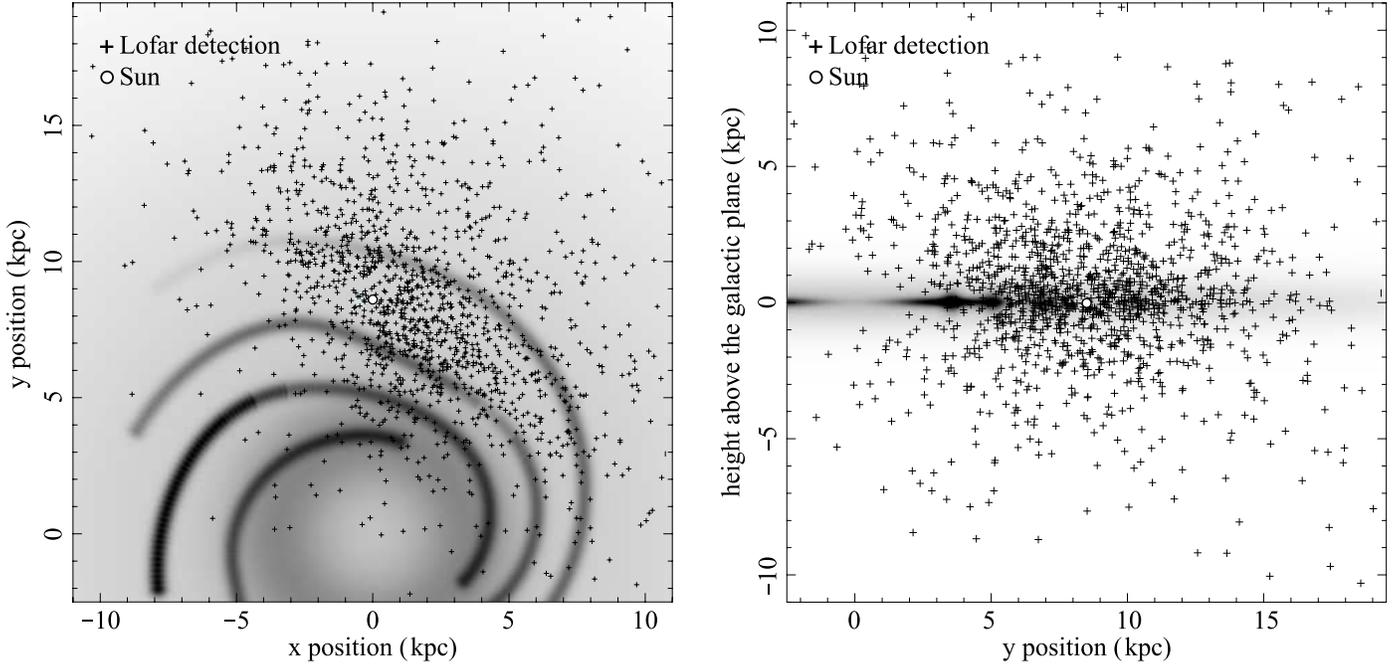}
	  \caption{The 1100 simulated pulsars detected
  with the reference Galactic LOFAR survey, for 1-hour pointings. The scattering
  interstellar matter (ISM) is shown in gray. {\bf Left)} projected on the
  Galactic plane. {\bf Right)} projected on the plane through the Galactic
  centre and sun, perpendicular to the disk.  The Galactic centre is
  at (0,0,0). LOFAR probes the local
  population very well, while the broadening from multi-path
  scattering on the ISM limits what volume
  a low-frequency survey like LOFAR probes. \citep[ISM modeled after][]{tc93}.
  \label{img:x-y-z} }
\end{figure*}

As a further check on our model, the PMB simulations find 800 $\pm$ 40
pulsars, which compares reasonably well to the actual number of 987
non-recycled pulsars detected \citep[ATNF catalog per Jan 1,
  2009]{mhth05}.

For LOFAR, we have evaluated surveys for different telescope
parameters, total survey time and minimum sensitivity. Our reference
survey is a {\em Full Incoherent} survey using 54 stations, 64-minute
pointings, a gain per Eq. \ref{eq:smin} of 0.66 times the theoretical
gain, with 80\% of 32MHz bandwidth at a central frequency of
140\,MHz. With a single beam such a survey can scan the visible sky in
25 days. In an all-sky survey with this reference set up 1100 $\pm$
100 pulsars could be detected (the full curve in
Fig.\ \ref{img:n-vs-t}), 900 $\pm$ 100 of which are new. Our
sensitivity estimate B uses a less conservative sensitivity of
0.8 times the theoretical gain. In the same overall time it finds
about 100-200 more pulsars that are new; i.e. not detected by older
surveys we simulate.  We do not take into account findings from
ongoing surveys such as the GBT350 search project
\citep{hrk+07,asr+09}. Here, and below, the errors quoted for
simulation runs indicate variations in output caused by using
different initial random seeds. Between simulation runs, the number of
new detections is a stable fraction of the total number of detected
pulsars, indicating that in the simulations there is a well-defined
part of the pulsar population that only LOFAR is sensitive too. In
contrast, there are also known pulsars that LOFAR will not be able to
detect in this reference survey, mainly due to scattering effects.  We
have also evaluated pointings of 1,2,4--256 minutes duration (see
Fig.\ \ref{img:n-vs-t}). Here the duration effects the minimum
detectable flux; using the figures of merit in Table \ref{tab:fom}
this incoherent survey can be scaled to the other survey types.

If pulsar luminosity is the limiting factor in the detections, a
larger time per pointing t decreases the minimum detectable flux
$S_{min}$ as $t^{-\frac{1}{2}}$, increasing the distance d out to
which pulsars can be detected as $t^{\frac{1}{4}}$. At the scale of
the Galaxy pulsars are located in a disk; if one could observe pulsars
throughout the Galaxy, the number of detectable pulsars is $N \sim d^2 \sim t^{\frac{2}{4}}$. Locally, pulsars are distributed roughly
isotropically; then $N \sim d^3 \sim t^{\frac{3}{4}}$
\citep[cf.\ ][]{cor02b}. We find that for LOFAR surveys, N is even
less dependent on $t$ than $N \sim t^{\frac{1}{2}}$ because scatter
broadening is the limiting factor, not luminosity
(cf.\ Fig.\ \ref{img:n-vs-t}, where the $t^{\frac{1}{2}}$ line is
scaled to be equal to the reference sensitivity estimate at the
6.25\,day point).

Compared to the Parkes Multibeam survey, the pulsars found with LOFAR
are different in several ways. The PMB is more strongly focused on the
Galactic centre. The range of the population detected by LOFAR is
limited by scatter broadening.  Because of its higher sensitivity,
LOFAR detects more low-luminosity nearby pulsars
(Fig.\ \ref{img:x-y-z}a). A lower limit to pulsar luminosity of
1\,mJy\,kpc$^2$ at 400\,MHz has been suggested (\citealt{lml+98},
cf.\ similar lower limit at 1400\,MHz in \citealt{lfl+06}); with a
spectral index of $-$1.5, this compares to a luminosity at 120\,MHz of
6\,mJy\,kpc$^2$. If this lower limit exists the reference survey can
detect all pulsars that are beamed toward us up to 1\,kpc.  If it does
not exist, then this survey will certainly illuminate how many
low-luminosity pulsars were formed nearby \citep[cf.\ PSR~J0240+62
  in][]{hrk+07}: a number of importance if one wants to understand the
neutron-star birthrate.

Furthermore, LOFAR finds more pulsars out of the Galactic disk
(Fig.\ \ref{img:x-y-z}b) as its high sensitivity is unimpeded by
scattering in that direction. As pulsars are born in the plane (z=0),
this observed z-distribution could disclose their 
much debated birth velocity \citep[cf.][]{har97, acc02}.

For the detection of millisecond pulsars (MSPs) scatter broadening
will be the limiting factor.  Although there is much variation in
scattering measure for different lines of sight, on average the
scatter broadening at 120\,MHz is on the order of 1\,ms for sources
with DMs higher than 30\,pc/cm$^3$ (using the rough empirical
  $t_{sc}$\,=\,$ 2 \times 10^9 ~ \textrm{DM}^{3.8}$ relation between DM and
  scattering time by \citealt{shi94}) or distances of about 1\,kpc
\citep[using distance-scattering relation from][]{tc93}.  As many of
the currently known Galactic-disk MSPs are within this range of 1\,kpc
in which scatter broadening does not hinder detection
\citep[][]{mhth05}, LOFAR could probe the local population to a
much lower flux limit.

%------globular clusters--------------------
\begin{table}[hb]
  \centering
  \begin{tabular}%
%  {|l|r|r|r|r|r|r|r|}
  {lrrrrrrr}
  \hline
   Name     &$DEC$ &   $d$& DM                  & $r_c$   & $\Gamma$ & $N$\\
            & deg  &  kpc & pc/cm$^3$           & arcmin  &          &    \\
  \hline\hline
 M15        & +12  & 10.3 & 67.3\hspace{1.6mm}  & 0.07    &     665  &  8 \\
 M92        & +43  &  8.2 & 29$^+$              & 0.23    &     106  &    \\
 M5         & +02  &  7.5 & 30.0\hspace{1.6mm}  & 0.42    &     79   &  5 \\
 M10        & -04  &  4.4 & 43$^+$              & 0.86    &     28   &    \\ 
 M13        & +36  &  7.7 & 30.4\hspace{1.6mm}  & 0.78    &     39   &  5 \\ 
 M3         & +28  & 10.4 & 26.3\hspace{1.6mm}  & 0.55    &     66   &  4 \\
 M2         & -01  & 11.5 & 29$^+$              & 0.34    &    120   &    \\
 M14        & -03  &  9.3 & 76$^+$              & 0.83    &     54   &    \\
 Pal 2      & +31  & 27.6 & 97$^+$              & 0.24    &     209  &    \\
 M12        & -02  &  4.9 & 39$^+$              & 0.72    &     10   &    \\
  \hline
  \end{tabular}
  \caption{Name,  declination ($DEC$), distance ($d$), dispersion measure (DM) observed or
  $^+$modeled after \citet{tc93}, core radius ($r_c$), collision number
  ($\Gamma$) and number of detected pulsars ($N$) for the ten
  highest-ranking candidates for a LOFAR globular cluster survey. Data
  from \citet{har96} and \citet{fre04b}.}
% numbers updated may 2009
  \label{tab:gc}
\end{table}
%--------------------

\subsection{Galactic globular cluster surveys}

In a fixed-time all-sky survey, one can gain field of view at a cost
of instantaneous sensitivity by adding the signal of the stations
incoherently instead of coherently. When the incoherent FoM is higher
(as in Table \ref{tab:fom}) this trade-off increases the number of
detectable pulsars because it allows for more time per pointing. In
contrast, specific, smaller regions on the sky with higher densities
of radio pulsars can potentially be better targeted with smaller field
of view, but significantly higher sensitivity.

\begin{figure}[t]
  \includegraphics[]{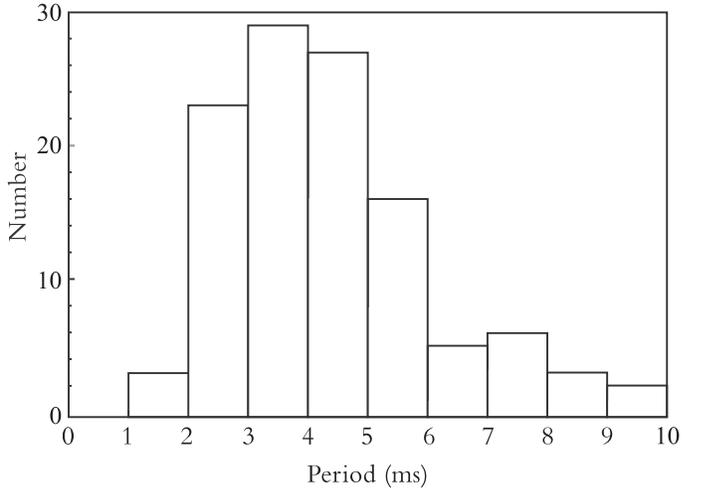}
  \caption{
	Periods of the 115 MSPs in globular
  clusters with periods less than 10\,ms. There are another 23
  pulsars with longer periods. \citep{fre04b}.
}
  \label{img:gc-ms-p}
\end{figure}
%----------------------------

%----------------------------
\begin{figure}[b]
  \includegraphics[]{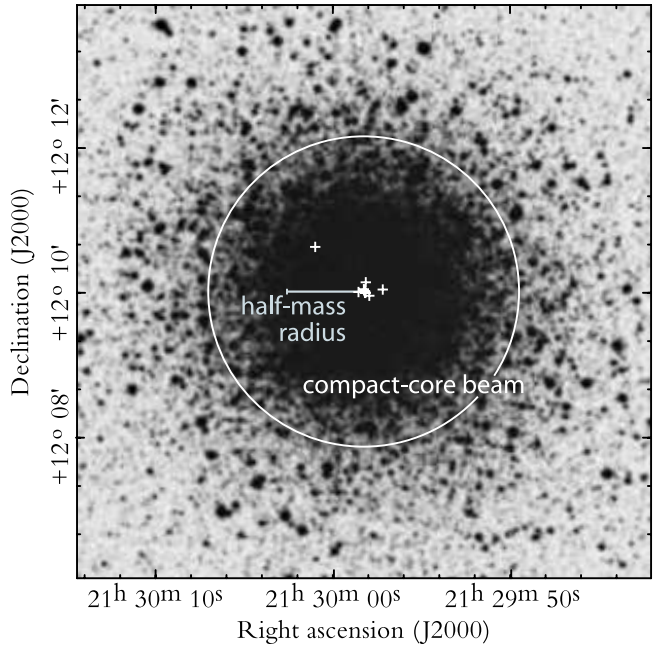}
  \caption{
	Beam setup for a MSP survey of
  M15, the globular cluster with the highest success probability in
  Table \ref{tab:gc}. The cluster half-mass radius falls
  well within the core beam. The
  crosses mark the 8 pulsars currently known \citep{and92}.
   }
  \label{img:m15}
\end{figure}
%----------------------------

Globular clusters fit the description well; they are compact and form
regions on the sky with high stellar densities. These high densities
also cause globular clusters to contain more binaries and binary
products than are found in the disk. This makes globular clusters very
good candidates for MSP searches \citep[e.g.\ ][]{lbm+87}. To estimate
which clusters are most promising for a LOFAR search, we evaluated
several of their properties: location on the sky, dispersion measure
(DM), distance ($d$), and the number of radio pulsars potentially
present.

As discussed above for sources in the Galactic plane, scatter
broadening is a concern for detecting far-away fast millisecond
pulsars. Most globular cluster pulsars have periods of around 2--5\,ms
\citep[][see Fig.\ \ref{img:gc-ms-p}]{fre04b}. Using the above
mentioned DM versus scattering time relations, at 120\,MHz such MSPs
are detectable up to a DM of about 30--40\,pc/cm$^3$. Observing at
200\,MHz extends this limit to 60--80\,pc/cm$^3$. For the longer
period pulsars also present in these clusters, these limits are less
of a problem, so we will further investigate globular clusters with
DMs up to 100\,pc/cm$^3$ below.  From studies with the Westerbork
telescope \citet{skh08} also conclude that the DM versus scattering
time relation is uncertain to the extent that some $DM=100\,$pc/cm$^3$
MSPs may be detectable.  We rate the candidate clusters by the
expected number of detectable pulsars: assuming $\frac{d~ log N}{d~
  log L}=-$1 \citep{and92}, this number scales as d$^{-2}$, as the
declination-dependent telescope gain $G(z)$ (cf. Figure
\ref{img:a_eff}), and as the collision number
$\Gamma\equiv\rho_c^{3/2}r_c^2$, where $\rho_c$ and $r_c$ are the
central density and the core radius, respectively \citep{ver03}. The
sources that are most promising according to this scaling are listed
in Table \ref{tab:gc}. The half-mass radius of each fits within a
LOFAR coherent core beam (cf. Fig.\ \ref{img:m15} for highest-ranking
candidate M15), which means we can now use the full gain of the
coherent addition.

In general, searching globular clusters more deeply than done
previously requires integrating for several hours and analyzing the
data with multiple binary-acceleration search techniques
\citep[see][and references therein]{rhs+05}.
Assuming an MSP spectral index of $-1.7$ \citep{kxl+98} the
weakest pulsar in M15 \citep[B2127+11G, 0.13\,mJy at 430\,MHz,
  from][]{and92} is 1.1\,mJy at 140\,MHz; in a 10-hour pointing with the
compact core at 140\,MHz, the minimum detectable flux
(eq.\ \ref{eq:smin} at M15's zenith angle of 40$^\circ$) is
0.5\,mJy. Thus far, the luminosities of the pulsars in M15 follow
$\frac{d~log N}{d~log L}=-$1. If this relation extends to lower
luminosities, observing up to a minimum flux of 0.5\,mJy could yield
several new MSPs.

\subsection{Pulsars in other galaxies}

The next over-dense regions on the sky are galaxies. Their
distance is a problem, but compared to globular clusters, galaxies
have several advantages for a LOFAR pulsar survey. They are
distributed equally over both hemispheres, while most globular
clusters are invisible from the telescope site in The Netherlands; and if visible face-on and
located in the part of the sky that is pointed away from our Galactic
disk, the scatter broadening is relatively low. 

%------galaxies-----------------------------
\begin{table}[b]
  \centering
  \begin{tabular}%
%  {|l|r|r|r|r|r|r|}
  {lrrrrrr}
  \hline
   Name     &   $d$&         $s$ & $i$   &   $gb$    & $\log M$    \\
            &  Mpc &     arcmin  & deg   &    deg    &  M$_\odot$  \\
  \hline\hline
 M31     &   0.7 &   193  &   78   &  -22   &  11.4  \\
 M81     &   1.4 &    22  &   60   &   41   &  10.7  \\
 M33     &   0.7 &    56  &   56   &  -31   &  10.1  \\
 M94     &   4.3 &    12  &   33   &   76   &  10.8  \\
 NGC2403 &   4.2 &    23  &   62   &   29   &  10.7  \\
 NGC4236 &   2.2 &    19  &   73   &   47   &   9.8  \\
 NGC4244 &   3.1 &    15  &   90   &   77   &  10.0  \\
 NGC4395 &   3.6 &    13  &   38   &   82   &  10.1  \\
 NGC3077 &   2.1 &     5  &   43   &   42   &   9.1  \\
 UGC7321 &   3.8 &     5  &   90   &   81   &   9.7  \\
  \hline
  \end{tabular} 
  \caption{Name, distance ($d$), size on the sky ($s$), inclination
    angle ($i$), Galactic latitude ($gb$) and logarithm of the mass
    ($\log M$) for the ten highest-ranking candidates for a LOFAR
    extragalactic pulsar survey. Data taken from \citet{tul88}.  }
  \label{tab:egal}
\end{table}

In the relatively nearby galaxy M31, a periodicity search based on a
10 hour pointing with the compact core at 140\,MHz could detect all
pulsars more luminous than $\sim$200\,Jy\,kpc$^2$ (see
eq.\ \ref{eq:smin}), which is comparable to the top end of the
luminosity distribution in our own galaxy. Any M31 pulsars that emit
{\em giant} pulses \citep{ag72} could be discovered more easily
through these giant pulses than through their periodicity if the flux
ratio between giant and normal pulses exceeds 10$^5$ \citep{mc03}. Two
young pulsars, the Crab pulsar B0531+21 and LMC pulsar B0540-69
\citep{sr68,jr03}, are known to emit such giant pulses and the former
was indeed discovered through them.  Compared to the regular periodic
emission, giant pulses show a steeper spectral index \citep[$-3.0$ to
  $-4.5$, ][]{vou01,mc03} making giant-pulse searches especially
attractive for a low-frequency telescope like LOFAR.  Using the
\citet{mc03} maximum distance estimator for giant-pulse detection in
1-hr pointings
\begin{equation}
 d = 0.85\,\textrm{Mpc} ~ \left(\frac{5\,\textrm{Jy}}{S_{sys}} \right) ~ 
\left(\frac{S_{GP}}{10^5\,\textrm{Jy}} \right)^{1/2} ~
\left(\frac{B}{10\,\textrm{MHz}}\right)^{1/4} 
\label{eq:dGP}
\end{equation}
with
$S_{sys}\,$=$\,G_{core}/T_{sys}$\,=\,$(0.8 \times 8.8\,$K/Jy$) / (1.0 \times 10^3\,$K$)$\,=\,$140\,$Jy,
bandwidth $B$\,=\,$32\,$MHz and scaling the giant pulse flux with
spectral index $-3.0$, the Crab pulsar would be detectable out to
$\sim$1.5\,Mpc. 

Detection of a handful of pulsars in each of several nearby galaxies
would enable comparison of the top end of their luminosity functions,
and potentially show a relation with galaxy type. Through pulse
scattering and dispersion these pulsars will also probe intergalactic
matter.

\begin{figure}[]
  \includegraphics[]{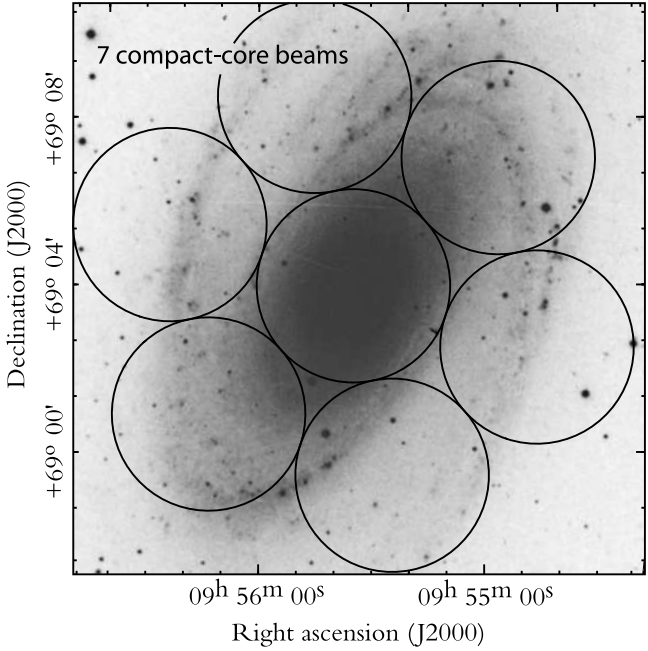}
  \caption{
	Seven coherently formed beams from the 2-km LOFAR core
        projected on candidate galaxy M81 at their FWHM.
   }
  \label{img:m81}
\end{figure}

Like for the globular clusters above we have ranked target galaxies by
distance, size on the sky and expected number of sources (here assumed
to scale linearly with mass). We select galaxies that are within
5\,Mpc from the Earth and that are not too strongly scatter broadened
given their Galactic latitude $gb$ and inclination angle $i$
(requiring $|gb|$+$90$-$i > 30$). In Table \ref{tab:egal} we list the
10 highest-ranked candidates.  For most of these highest-ranked
galaxies the core can be captured with a single compact-core beam
(cf. Fig.\ \ref{img:m81}), or the whole galaxy by a pattern of 7
beams.

%----------------------------------------------------
\section{Discussion}
\label{sec:disc}

\subsection{Steep spectrum sources}
Generally pulsars are increasingly bright toward lower frequencies,
but this steep spectrum of pulsars typically flattens out or turns over
at a frequency between 100 and 250\,MHz \citep{mgj+94}. There is,
however, a small fraction of pulsars for which no such spectral break
is observed down to frequencies as low as 50\,MHz. Sources with a
spectral index steeper than the spectral index of the sky background
of $-$2.6 \citep[e.g.\ PSR~B0943+10, spectral index $-$4.0; ][]{rd94}
will be more easily detectable by a telescope like LOFAR, producing
quantitative input for radio emission models.

\subsection{Follow up}
With high sensitivity most important and field of view not a
consideration, follow up timing \citep[the determination of pulsar and
  potential binary parameters, cf.][]{lk05} will use the entire
core, and potentially more of the LOFAR array if it can be reliably
phased up, for beam forming. Given the small $3'$ field of view, such
follow up needs very accurate candidate source positions. Determining
the source position that accurately also directly facilitates timing
follow-up by eliminating position as a variable.

In some of the above surveys sources are only localised to within the
$30'$ superstation or $5^\circ.8$ station beams (cf. Section
\ref{sec:bf}). 
% those are sizes at 120MHz.
Detections can be more accurately located later by
following up with many, smaller tied-array beams. As the DM is now
known the total computational burden of searching through these many
tied-array beams is not as problematic as for an entirely coherent
survey. For example, we find that about 50\% of incoherent-survey
pointings will have sources brighter than the reference
$S_{min}$. Each of these station-beams pointings could subsequently be
filled out with 128 coherent superstation beams in about 1/3rd of the
original total survey time ($50\% \,
(\frac{FoM_{Full}}{FoM_{Superstation}})^2\,=\,0.36$, cf. Eq. \ref{eq:fom}
and Table \ref{tab:fom}) to reach the same $S_{min}$. This would
provide the necessary first conformation observation but also an improved
position. In a next step 128 compact-core beams can then similarly tile out the
superstation beam in which the source is located to
provide full localisation, and the first timing data.

Once sources are properly localised, long-term follow up timing uses
tied-array beams. From our simulations we find that in the incoherent
reference survey about 80\% of sources have nearest neighbors at less
than one FWHM station beam away and hence these pairs of sources can
be timed simultaneously. 18\% and 12\% of simulated pulsars are in
groups of 3 and 4 per station field-of-view respectively. In total the
follow up needs to re-observe about 50\% of the survey pointings to
time each newly found pulsar once. Sub-arraying or trading bandwidth
for independently steerable station beams could potentially increase
overall follow-up timing efficiency. The timing observation should
hold at least several hundred pulses for a steady profile to form,
which at an average 1\,s-pulsar translates to about 10\,min
integrations.  Given the $\sqrt{N_{stations}}$\,=\,$7.3$-fold increase
in gain between incoherent surveying and tied-array follow-up, a
timing run can thus produce high SNR detections for all candidates.

\subsection{Fast radio transients}
Given the software-telescope nature of LOFAR, piggybacking and
simultaneous observing are relatively straightforward to implement. As
the 'telescope time' concept at LOFAR includes the availability of
central signal processing at CEP, observation modes that are not
computationally or IO intensive, such as the incoherent station
addition mode, are especially suited for parallel observing. This
incoherent mode can therefore run commensally with many of the imaging
projects \citep{rbb+06,fws+08} to continuously scan a large part
of the sky for intermittent pulsars such as PSR~B1931+24
\citep{klo+06} and millisecond radio transients \citep{lbm+07,hsl+09}.

\subsection{Survey strategies}
Three types of surveys were presented and their overall efficiency
(Table \ref{tab:fom}) is comparable. When deciding which LOFAR survey
to implement, considerations relating to the different integration
times, beam sizes and operational requirements in these surveys may
thus become more important: {\em Core Coherent} has short (order
$\sim$1\,min) integrations, beneficial for finding accelerated systems
and for system stability, and provides good localisation. {\em
  Superstation Coherent} and {\em Full Incoherent} use of order
$\sim$1\,hr integrations, and are well suited for finding intermittent
pulsars, but with only reasonable and poor initial localisation
respectively. {\em Core Coherent} and {\em Superstation Coherent}
produce much higher data rates than {\em Full Incoherent}. Tests on
the different types of LOFAR data that are currently becoming
available may potentially further differentiate between these
strategies.

\section{Conclusions}
Because of its large area and field of view, LOFAR can reveal the local
population of pulsars to a very deep luminosity limit. A 25-day
all-sky survey at 140\,MHz would find 900 new pulsars, disclosing the
local low-luminosity population and roughly doubling the number of
pulsars known in the northern hemisphere. Millisecond pulsars in
nearby globular clusters can be detected to lower flux limits than
previously possible. Assuming the pulsar population in other galaxies
is similar to that in ours, we can detect periodicities or giant
pulses from extragalactic pulsars up to several Mpcs away.

\bibliographystyle{aa}
\bibliography{journals_apj,modrefs,psrrefs,crossrefs,modjoeri}

\end{document}